\documentclass{iau}
\usepackage{graphicx}

\title[The Nuclear Cluster of the Milky Way] 
{The Nuclear Cluster of the Milky Way: Total Mass and Luminosity}

\author[T. K. Fritz]   
{T. K. Fritz$^{1,2}$,
  S.~Chatzopoulos$^1$, O. Gerhard$^1$
, S.~Gillessen$^1$,  R.~Genzel$^{1,3}$, 
O.~Pfuhl$^1$, 
 S.~Tacchella$^4$, F.~Eisenhauer$^1$, T.~Ott$^1$
}

\affiliation{$^1$Max Planck Institut f{\"u}r Extraterrestrische Physik, Postfach 1312, D-85741, Garching, Germany 
\\[\affilskip]
$^2$ Department of Astronomy, University of Virginia, 530 McCormick Road, Charlottesville, VA 22904-4325, email: {\tt tkf4w@virginia.edu}
\\[\affilskip]
$^3$ Department of Physics, University of California, Berkeley, 366 Le Comte Hall, Berkeley, CA 94720-7300
\\[\affilskip]
$^4$Institute for Astronomy, ETH Zurich, Wolfgang-Pauli-Strasse 27, CH-8093, Zurich, Switzerland

}

\pubyear{2013} 
\volume{303}  
\pagerange{}
\jname{The Galactic Center: Feeding and Feedback in a Normal Galactic Nucleus}
\editors{Editors}
\begin{document}

\maketitle

\begin{abstract}

 Here we present
the fundamental properties of the nuclear cluster of the Milky Way.
First, we derive its structural properties by constructing a density map of the central 1000'' using extinction-corrected 
star counts. We can describe the data with a two-component model built from Sersic profiles. The inner nearly spherical component
is the nuclear cluster.
The outer, strongly flattened component can be identified with the stellar component of the circumnuclear zone.
Second, we enlarge the radius inside which detailed dynamics are available from $1\,$pc to $4\,$pc.
We use more than 10000 individual proper motions and more than 2700 radial velocities.
We determine the cluster mass by means of isotropic spherical Jeans modeling.
We get a nuclear cluster mass within 100'' of $M_{100''}=(6.11 \pm 0.52|_{\mathrm{fix} R_0}\pm 0.97|_{R_0}) \times 10^6 $ M$_{\odot}$, 
which corresponds to a total cluster
mass of  M$_{\mathrm{NC}}=(13.08 \pm 2.51|_{\mathrm{fix} R_0}\pm 2.08|_{R_0}) \times 10^6 $ M$_{\odot}$.
By combination of our mass with the flux  we calculate $M/L=0.50 \pm 0.12  M_{\odot}/L_{\odot,\mathrm{Ks}}$  for the central 100''. 
That is broadly consistent with a Chabrier IMF. 
With its mass and a luminosity of
M$_{\mathrm{Ks}}=-15.30\pm0.26$ the nuclear cluster is a bright and massive specimen with a typical size.
\end{abstract}

\firstsection 
\section{Introduction}

In the centers of many late-type galaxies one finds massive stellar clusters: the nuclear star clusters 
(\cite[Phillips et al. 1996]{Phillips_96}, \cite[Matthews 
\& Gallagher 1997]{Matthews_97}, \cite[Carollo et al. 2002]{Carollo_02}). 
The nuclear clusters are central light overdensities on a scale of about 5~pc (\cite[B{\"o}ker et al. 2004]{Boeker_04}). 
Also, the central light concentration 
of the Milky Way (\cite[Becklin \& Neugebauer 1968]{Becklin_68}) is a nuclear star cluster (\cite[Philipp et 
al. 1999]{Philipp_99}, \cite[Launhardt et 
al. 2002]{Launhardt_02}).
Nuclear clusters are comparable in density to globular clusters, but are typically more massive (\cite[Walcher et al. 2005]{Walcher_05}).

Due to the proximity of the Galactic Center (GC), the nuclear cluster of the Milky 
Way can be observed in much higher detail than any other nuclear cluster (\cite[Genzel et al. 2010]{Genzel_10}). It is useful to shed light on the 
properties of nuclear clusters and their formation in general. Here and in Fritz et al. submitted we concentrate on the most basic properties, mass and luminosity. 
 For that aim we obtain the star distribution of the central galaxy out to 1000'', which allows us to cover far more than the nuclear cluster. In 
addition, we measure for the first time motions in all three dimensions over a large part of the cluster expanding on
\cite[Lindqvist et al. (1992)]{Lindqvist_92a}, \cite[Genzel et al. (1996)]{Genzel_96}, \cite[Trippe et al. (2008)]{Trippe_08} and \cite[Sch{\"o}del et 
al. (2009)]{Schoedel_09}.

\section{Density Profile and Luminosity}
 \label{sec2}

 For the density maps we combine (J/H/)Ks-band data from NACO, WFC3/IR and VISTA. The NACO map provides the high resolution data in the center while the VISTA map gives the largest field of view. With this combination, we have sufficient resolution nearly everywhere for azimuthal decomposition.  We use the data in two 
different ways: first, for number counts of bright stars; 
and second, for the integrated flux. 
All data are corrected for extinction by using two color information and masking out infrared dark clouds.
We also remove the contribution of the young (OB and IRS7) stars in order to exclude the most recent starburst, whose radial profile is different
from the older stars (\cite[Bartko et al. 2010]{Bartko_10}, \cite[Blum et al. 2003]{Blum_03}, \cite[Pfuhl et al. 2011]{Pfuhl_11}). 

\begin{figure}[b]
\begin{center}
 \includegraphics[width=5.32in]{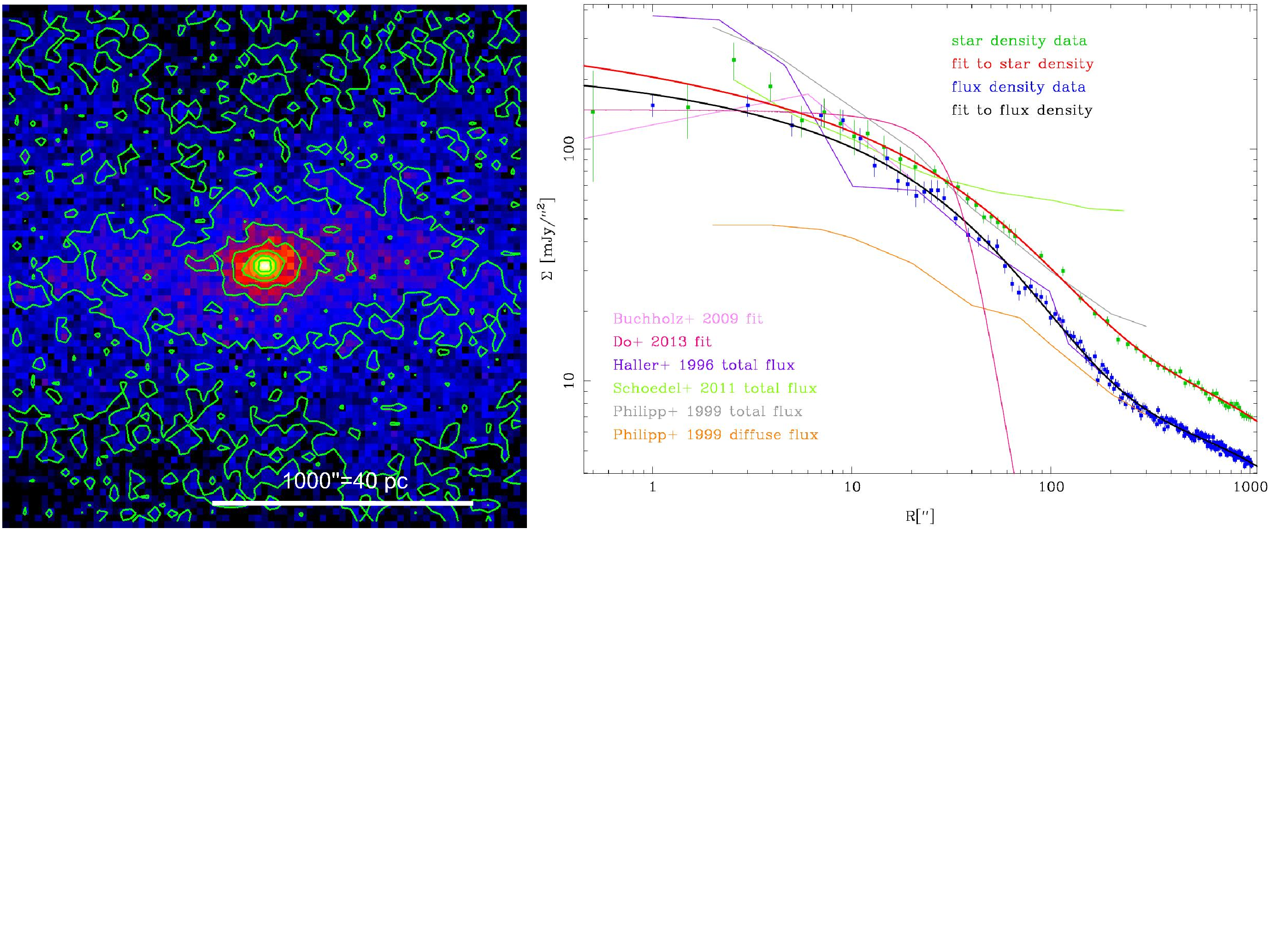} 
 \caption{
Star distribution in the GC.  {\it Left panel}: Star density map around Sgr~A* (r$=1000$'' box) with smoothed contours.  {\it Right panel}: Profiles from our data in 
comparison with the literature. All curves besides our flux density haven been scaled, since they do not have the same units or are not extinction-corrected. 
The thick lines show our fits with each consisting of two independent $\gamma$-models.
}
   \label{fig2}
\end{center}
\end{figure}

The resulting map is presented in Fig.\,\ref{fig2}.
It is apparent that the star distribution is close to circular in the center and more elongated further out. This strengthens the finding of \cite[Launhardt et 
al. (2002)]{Launhardt_02} that the star-forming nucleus of the Milky Way consists of two components, one which is the close to circular nuclear cluster and a second, which is a flattened nuclear disk. The 
latter is the stellar component of the circumnuclear disk.
To differentiate between these components, we fit them with two Sersic profiles (using GALFIT). We obtain for the nuclear cluster an axis ratio of $1.10\pm0.07$,  
a half light radius of $110 \pm 10''=4.4\,$pc, a Sersic index of $1.42\pm0.03$ and a total luminosity of M$_{\mathrm{Ks}}=-15.3\pm 0.26$. The main uncertainties
include the modeling of the incompletely-covered outer component as well as the extinction correction.

We present the flux and number counts radial profile in Fig.\,\ref{fig2}. They are not in perfect agreement with one another, yet, there is similarity between the two and they appear to agree roughly with \cite[Becklin \& Neugebaur (1968)]{Becklin_68}, \cite[Haller et al. (1996)]{Haller_96} and \cite[Philipp et 
al. (1999)]{Philipp_98}. 
We as well as the aforementioned studies seem to be in disagreement with the recent profile of \cite[Sch{\"o}del (2011)]{Schoedel_11},
who measured a larger background flux level possible due to missing sky subtraction.
Our profile also deviates strongly from the
indirect, dynamic profile determination of \cite[Do et al. 2013]{Do_13}.  In comparison to ours, their profile has a too large core and a too steep of an outside decline.
Note that we fit our profiles for the mass modeling with a combination of two spherical $\gamma$-models (\cite[Dehnen 1993]{Dehnen_93}).
By using both, flux and number counts, we include density profile uncertainties in our error budgets.

\section{Mass}
 \label{sec3}

Additionally, to obtain the mass, we measure star velocities in all three dimensions. We employ mainly the NACO data to generate more than 10000 proper motions out
to 80''. For radial velocities, we use SINFONI to get more than 2500 radial velocities out to 80'' and to go beyond this region, we add about 200 maser velocities 
from \cite[Lindqvist et 
al. (1992)]{Lindqvist_92a}  and \cite[Deguchi et al. (2004)]{Deguchi_04}. To remove the young stars from our velocity-selected sample, we employ spectra. 
  
Our data show no sign of radial anisotropy out to at least 40'', in contrast to the extrapolation of \cite[Do et al. (2013)]{Do_13}.
Beyond this, the azimuthal coverage is too incomplete to distinguish radial anisotropy from other deviations from isotropy like flattening. 
The difference between the dispersion in and perpendicular to the galactic plane (\cite[Trippe et al. 2008]{Trippe_08}) is explained by flattening of 
the nuclear cluster in that direction (see Chatzopoulos et al., in prep). We confirm the rotation of the cluster via our
radial velocity measurements. The rotation velocity 
is, however, on the small side of \cite[McGinn et al. (1989)]{McGinn_89} and \cite[Trippe et al. (2008)]{Trippe_08} outside of 1 pc.

\begin{figure}[b]
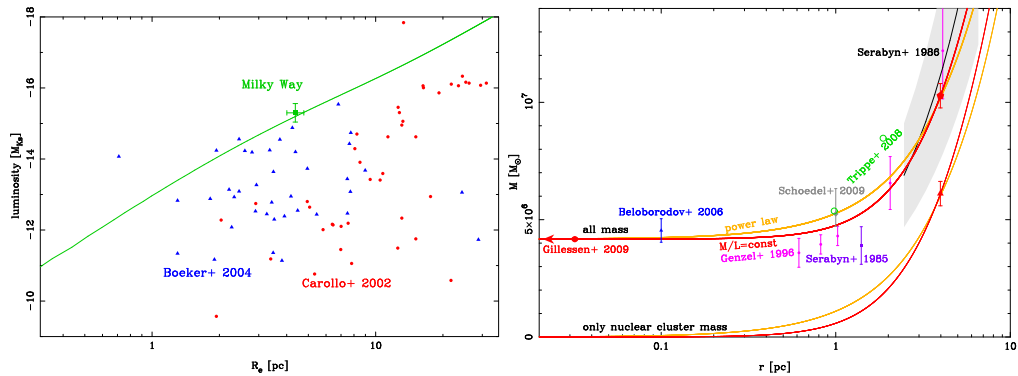

\begin{center}
\includegraphics[width=1.92in,angle=-90]{nc.par1.eps} 
\includegraphics[width=1.92in,angle=-90]{tot.mass.s-f.eps} 
 \caption{{\it Left panel}: Cumulative mass distribution in the GC (with and without the SMBH) in comparison with the literature.
The red and orange lines show the results of this work.
{\it Right panel}: The GC in comparison with \cite[Carollo et al. (2002)]{Carollo_02} and \cite[B{\"o}ker et al. (2004)]{Boeker_04}. In case of the GC, we show both
the size of the nuclear cluster which follows from our two dimensional decomposition (square) and the cumulative flux curve.
}
   \label{fig3}
\end{center}
\end{figure}

We obtain the mass by Jeans modeling. Here, we restrict ourselves to isotropic spherical symmetric modeling. The mass is modeled by the supermassive black 
hole (SMBH) in the center and the extended nuclear cluster mass. The SMBH mass is fixed to the black hole distance relation
of \cite[Gillessen et al. (2009)]{Gillessen_09}. Since the exclusion of flattening in the model a statistical parallax distance would lead to biasing, we
use the independent measurement of \cite[Gillessen et al. 2013]{Gillessen_13} (R$_0=8.2\pm0.34\,$kpc). For the nuclear cluster we use two models: a power law
 mass profile and a constant mass to light ratio. We obtain a nuclear cluster mass of $(6.11 \pm 0.52|_{\mathrm{fix} R_0}\pm 
0.97|_{R_0} ) \times 10^6  M_{\odot}$ within 100'' which is consistent with most of the literature (Fig.\,\ref{fig3}). The distance-independent mass error is dominated by 
the profile difference between star counts and 
light density. The error does not include deviations from our model like flattening. Using the two-dimensional decomposition of Sec.\,\ref{sec2} 
we obtain a total nuclear cluster mass of $(13.08 \pm 2.51|_{\mathrm{fix} R_0}\pm 2.08|_{R_0}) \times 10^6  M_{\odot}$.

Employing both the mass and luminosity, we find a mass to light ratio of $M/L=0.50 \pm 0.12 \, M_{\odot}/L_{\odot,\mathrm{Ks}}$. 
  By combining the GC star formation history 
(\cite[Pfuhl et al. 2011]{Pfuhl_11}, \cite[Blum et al. 2003]{Blum_03})
with the canonical Kroupa/Chabrier IMF, we obtain $M/L\approx0.68$, as expected value for the cluster. The simplicity of our mass model could possibly cause 
differences/errors.

\section{Comparison with other Nuclear Clusters and Origin}
 \label{sec4}

 To compare the cluster of the Milky Way with other nuclear clusters, we use the size and luminosity measurements of 
 \cite[Carollo et al. (2002)]{Carollo_02} and \cite[B{\"o}ker et al. (2004)]{Boeker_04}, which cover mainly early- and late-type spirals, respectively.
The size of the cluster is in between these two samples as expected from the Hubble type of the Milky Way. The cluster is however brighter than most other nuclear
clusters. That result holds also for the cumulative flux curve which is unaffected by decomposition uncertainties. 
 Our measurement of an usual M/L shows that the cluster of Milky Way has also an unusually high mass. Since its size is typical, this implies that the Milky Way has a higher star
density than most nuclear clusters.

The bright nuclear cluster of the Milky Way is like most bright nuclear clusters 
associated with a bright active circumnuclear region (\cite[Launhardt et al. 2002]{Launhardt_02}, \cite[Carollo et al. 1999]{Carollo_99}).
Seeing that the majority stars in the nuclear cluster are old (\cite[Pfuhl et al. 2011]{Pfuhl_11}), 
this makes it likely also that its old stars have more connections with the nuclear disk than with globular clusters (\cite[Tremaine et al. 1975]{Tremaine_75}). It is 
however unclear whether
a nuclear disk origin of the nuclear cluster can be reconciled easily with a difference in the flattening in the two components (Sec.\,\ref{sec2}).

\end{document}